\documentclass[a4paper,fleqn,usenatbib]{mnras}

\usepackage{txfonts}

\usepackage[T1]{fontenc}
\usepackage{ae,aecompl}

\usepackage{color}

\usepackage{graphicx}
\usepackage{amstext}

\usepackage{hyperref}
\usepackage{bookmark}

\newcommand{\rxte}{\emph{RXTE}}
\newcommand{\chandra}{\emph{CHANDRA}}
\newcommand{\xmm}{\emph{XMM-NEWTON}}
\newcommand{\SF}{GRO~J1655$-$40}
\newcommand{\wsim}{\ensuremath{\sim}}

\title[GRO~J1655$-$40 hard flares]{Wind, jet, hybrid corona and hard X-ray flares: multiwavelength evolution of GRO~J1655$-$40 during the 2005 outburst rise}

\author[E. Kalemci et al.]{
E. Kalemci$^{1}$\thanks{E-mail:ekalemci@sabanciuniv.edu}, M.~C. Begelman$^{2,3}$, T.~J. Maccarone$^{4}$, T. Din\c{c}er$^{5}$, T.~D. Russell$^{6}$
\newauthor C. Bailyn$^{5,7}$, J.~A. Tomsick$^{8}$
\\
$^{1}${Faculty of Engineering and Natural Sciences, Sabanc\i\ University, Orhanl\i-Tuzla, 34956, Istanbul, Turkey}\\
$^{2}${JILA, University of Colorado and NIST, 440 UCB, Boulder, CO 80309-0440, USA}\\
$^{3}${Department of Astrophysical and Planetary Sciences, University of Colorado, Boulder, CO 80309, USA}\\
$^{4}${Department of Physics and Astronomy, Texas Tech University, Box 41051, Lubbock, TX 79409-1051, USA}\\
$^{5}${Department of Astronomy, Yale University, PO Box 208101, New Haven, CT 06520-8101, USA}\\
$^{6}${ International Centre for Radio Astronomy Research - Curtin University, GPO Box U1987, Perth, WA 6845, Australia}\\
$^{7}${Yale-NUS College, 6 College Avenue, East, , 138614, Singapore}\\
$^{8}${Space Sciences Laboratory, 7 Gauss Way, University of California, Berkeley, CA, 94720-7450, USA}
}

\date{Accepted XXX. Received YYY; in original form ZZZ}

\pubyear{2016}

\begin{document}
\label{firstpage}
\pagerange{\pageref{firstpage}--\pageref{lastpage}}
\maketitle

%\begin{document}
%\date{submitted to mnras}
%\pagerange{\pageref{firstpage}--\pageref{lastpage}} \pubyear{2014}
%\maketitle

%% /*******************************************************************
%% ** The Abstract                                                   **
%% *******************************************************************/

\begin{abstract}

We have investigated the complex multiwavelength evolution of \SF\ during the rise of its 2005 outburst. We detected two hard X-ray flares, the first one during the transition from the soft state to the ultra-soft state, and the second one in the ultra-soft state. The first X-ray flare coincided with an optically thin radio flare. We also observed a hint of increased radio emission during the second X-ray flare. To explain the hard flares without invoking a secondary emission component, we fit the entire data set with the \emph{eqpair} model. This single, hybrid Comptonization model sufficiently fits the data even during the hard X-ray flares if we allow reflection fractions greater than unity. In this case, the hard X-ray flares correspond to a Comptonizing corona dominated by non-thermal electrons. The fits also require absorption features in the soft and ultra-soft state which are likely due to a wind. In this work we show that the wind and the optically thin radio flare co-exist. Finally, we have also investigated the radio to optical spectral energy distribution, tracking the radio spectral evolution through the quenching of the compact jet and rise of the optically thin flare, and interpreted all data using state transition models.
 
\end{abstract}


%] %twocolumn

%% /*******************************************************************
%% ** Introduction                                                   **
%% *******************************************************************/

\section{Introduction}\label{sec:intro}

Galactic black hole transients (GBHT) are systems that occasionally go into outburst, during which their X-ray luminosity may increase several orders of magnitude when compared to their quiescent levels. These objects are excellent laboratories to study the complex relationship between jets, winds and the accretion environment as the outbursts evolve on time scales of months. This rapid evolution allows for the detailed investigation of the properties of accretion states which are traced by X-ray spectral and timing properties, and the properties of outflows, where the jets are traced by the radio and optical/infrared (OIR) emission and the winds are traced by the properties of X-ray and optical absorption features. 

A detailed description of X-ray spectral and timing states of GBHTs can be found in \cite{McClintock06book} and \cite{Belloni10_jp}. At the start of a typical outburst, the GBHT is in the hard state (HS). In this state, the X-ray spectrum is dominated by a hard, power-law like component associated with Compton scattering of soft photons by a hot electron corona. Faint emission from a cool, optically-thick, geometrically-thin accretion disc may also be observed, which can be modelled by a multi-temperature blackbody \citep{Makishima86}. This state also exhibits strong X-ray variability (typically $>$20\% of the fractional rms amplitude). As the outburst continues and the X-ray flux increases, the GBHT usually transitions to a soft state (SS) in which the X-ray spectrum is now dominated by the optically-thick accretion disc, displaying low levels of X-ray variability ($<$ a few \%) and faint power-law emission. Between the HS and the SS, the source may transition through the hard and soft intermediate states (HIMS and SIMS, respectively) with properties in between the hard and soft states \citep[see][for further details]{Belloni10_jp}. Finally, some sources (e.g. \SF, Cyg X-3) may also show a so-called ultra-soft state (US), which has an extremely steep X-ray power-law (with a photon spectral index $\Gamma$ of $>$3) with a completely dominating disc contribution \citep{Szostek08, Zdziarski04}. \SF\ shows an even softer X-ray state, denoted the ``hyper-soft" state \citep{Uttley15}.

GBHTs also show distinct multiwavelength characteristics throughout their outbursts. Radio and optical-infrared (OIR) observations indicate the presence of compact jets which exhibit flat to inverted radio spectrum (such that the radio spectral index $\alpha\gtrsim0$, $S_{\nu} \propto \nu^{\alpha}$ where $S_{\nu}$ is the radio flux density at frequency $\nu$) in the hard state \citep{Tananbaum72, Buxton04, Corbel13, Gallo10} which become quenched in the soft state \citep{Russell11, Fender99, Coriat11}. During the transition from the hard to soft state, the compact jets give way to relativistic and bright transient jets with an optically-thin radio spectrum \citep[$\alpha < 0$,][]{Fender09, Vadawale03}. Recent high-resolution grating observations of GBHTs and neutron stars revealed the presence of blue-shifted absorption features, especially \ion{Fe}{xxv} and  \ion{Fe}{xxvi} lines, showing that these sources not only produce collimated jets, but can also drive winds \citep[][and references therein]{DiazTrigo14, Neilsen12}. Wind signatures are preferentially detected in soft states for high-inclination sources \citep{Done07, Ponti12}, where the inclination dependence indicates a thermal origin for the wind \citep{Begelman83}. A single observation of \SF\ on MJD~53461.5 revealed a rich series of absorption lines from a dense, highly-ionized wind, which was initially interpreted as magnetically-driven \citep{Miller06}. Most follow up studies have supported the magnetic origin of the wind \citep[][and references therein]{Neilsen12}. However, thermally-driven winds remain a possibility \citep{Netzer06}.

\subsection{\SF}
\label{sub:source}

\SF\ was first discovered with the Burst and Transient Source Experiment (BATSE) on-board the Compton Gamma Ray Observatory \citep{Zhang94_iauc}. Subsequent radio observations revealed apparent-superluminal relativistic (0.92 $c$) jets \citep{Hjellming95, Tingay95}. Optical observations taken in quiescence indicate a FIII-FV giant or sub-giant with an orbital period of 2.62 days \citep{Orosz97}. In this study, we used primary and secondary masses of 6.3$\pm$0.5 $\rm M_{\odot}$ and 2.4$\pm$0.4 $\rm M_{\odot}$, respectively, which were obtained by modelling the ellipsoidal orbital modulations in quiescence \citep{Greene01}. The same model indicates a binary inclination of 70.2$^{\circ} \pm$1.9$^{\circ}$ which is consistent with deep absorption dips \citep{Kuulkers00} and strong wind emission \citep{Ponti12}.  Alternative mass measurements exist \citep{Beer02, Shahbaz03}, but the differences are small and have no effect on our conclusions. The binary inclination angle of \SF\ is slightly different from the disc inclination angle obtained from radio imaging \citep{Orosz97, Maccarone02}. The distance to the source has been estimated via different methods, where the majority of works use a distance of 3.2$\pm$0.2 kpc, based on the analysis of \cite{Hjellming95}, which we also adopt. 

Between 1994 and 1997 BATSE detected several outbursts from \SF\ \citep{Zhang97_2}. These early outbursts showed a complex pattern between the hard X-ray flares and the optically-thin radio flares;  the first three hard X-ray flares were very well correlated with superluminal radio flares \citep{Harmon95}. However, subsequent hard X-ray flares were not associated with any increased radio emission \citep{Tavani96, Zhang97_2}. 

This article investigates the multiwavelength evolution of \SF\ during its 2005 outburst rise, which was first detected on February 17 (MJD~53419) with the Proportional Counter Array (PCA) instrument on-board the Rossi X-ray Timing Explorer (\rxte) \citep{Markwardt05_atel}. The source was intensely monitored with \rxte\ throughout this outburst. There was also exceptional multiwavelength coverage during this outburst, which was followed daily in OIR with the Small and Moderate Aperture Research Telescope System (SMARTS; this work and \citealt{Kalemci13}), as well as frequent radio observations with the Very Large Array (VLA; this work and \citealt{Shaposhnikov07}). Grating observations with \chandra\ and \xmm\ taken during this outburst have also revealed wind features, the origin of which is still under debate \citep{Miller06, DiazTrigo14, Shidatsu16}. 

The structure of the paper is as follows. In \S\ref{sec:obs} we describe the multiwavelength observations and provide detailed information of the spectral extraction in all observing bands. In \S\ref{sub:states} we describe the source states and transitions during the rise, and then, for the first time, discuss the properties of the hard X-ray flares in \S\ref{sub:hard}. To explain the possible origin of hard X-ray flares, we conducted spectral fits with \emph{eqpair}, which are discussed in \S\ref{sub:eqpair}. The radio to OIR spectral energy distribution (SED) are shown in \S\ref{sub:SED}, with emphasis on the optically-thin radio flare. Finally we discuss our findings, focusing on the origin of hard X-ray flares and the relationship between the radio and wind emission.

%% /*******************************************************************
%% ** Observations and Analysis                                      **
%% *******************************************************************/

\section{Observations and Analysis}\label{sec:obs}

We have conducted a comprehensive analysis of X-ray, radio and OIR observations of \SF\ from the start of the 2005 outburst until the end of the ultra-soft state as the source entered the so called ``hyper-soft'' state. 

\subsection{X-ray observations and analysis}

We have analysed 46 \rxte\ observations between MJD~53422.9 and MJD~53461.6 utilizing both the PCA and the HEXTE instruments. These observations cover the initial HS, HIMS, SIMS, SS and most of the US. The \rxte\ \emph{obsids} are shown in Table~\ref{table:x1} in the appendix. The spectral extraction details can be found in \cite{Kalemci13}. We also conducted timing analysis to confirm the spectral states. Details of this procedure can be found in \cite{Dincer14}.

X-ray spectral fitting was done using two models: a phenomenological diskbb+power-law, and the more physical hybrid plasma Comptonization model \emph{eqpair} \citep{Coppi99}.  We employed an automatic fitting algorithm to determine the evolution of spectral parameters. For all observations in states other than the US, we started with the PCA spectrum and fitted them with a model that comprised interstellar absorption (\emph{tbabs} in XSPEC), power-law and a smeared edge (\emph{smedge}, \citealt{Ebisawa94}). We used cross-sections of \cite{Verner96} and abundances of \cite{Wilms00} for the interstellar absorption and fixed the  $N_{H}$ to $\rm 0.8 \times 10^{22} \, cm^{-2}$ \citep{Migliari07}. We added a multi-colour disc blackbody (\emph{diskbb} in XSPEC, \citealt{Makishima86}), testing its presence with an F-test.  We also tested for the presence of an iron emission line for each observation. We included a \emph{diskbb} and/or \emph{Gauss} component if the F-test chance probability was less that 0.005. We then added the HEXTE spectrum, leaving the normalization free and re-fitted the spectrum. We tested for the presence of a high energy cut-off by adding a \emph{highecut} to the overall model, including the cut-off component if the F-test chance probability was less than 0.005. Once we had obtained a reasonable spectral fit, we ran the error command for all free parameters to refine the fit and determine uncertainties.

For the US, the model used for the other states did not provide acceptable fits due to the presence of absorption features between 6 and 10 keV, which may be related to the iron absorption features (discussed in detail in \S\ref{sub:source} and \S\ref{sub:disOIR}). To account for these features, we first added a single Gaussian with negative normalization and constrained its energy between 6.2 keV and 7.2 keV. For some observations, we needed a second Gaussian with negative normalization between 7.5 and 8.5 keV. Assuming that they are related to the iron absorption features discussed in \cite{DiazTrigo07}, we restricted the line widths to be less than 0.5 keV.

For some observations, even after adding the Gaussian features, there were significant residuals in the HEXTE data at energies above 100 keV. To determine the flux of this residual emission, we added a second power-law and constrained its index to between 1.2 and 2, assuming this is a component which mostly affects the hard X-rays and does not alter the spectrum significantly below 20 keV. The results of the fits are tabulated in Table~\ref{table:x1}\footnote{All errors in the figures and in tables correspond to $\Delta\chi^{2}$ of 2.706.}.

We also fitted all observations with the \emph{eqpair} model. Since \emph{eqpair} has many parameters, leaving all of them free results in fits which are harder to interpret due to degeneracy between some parameters. For this reason, for each spectral state we employed the recipe provided in \cite{Coppi99}. The current version of \emph{eqpair} in XSPEC allows the user to choose between a blackbody or \emph{diskpn} \citep{Gierlinski99} as the soft photon input from the disc. We used \emph{diskpn} to keep the disc emission and the soft photon input consistent. Note that we have also fitted all observations with \emph{diskbb}+\emph{eqpair} with the blackbody temperature fixed to the \emph{diskbb} inner disc temperature (implying a patchy corona) and confirmed that the general evolution of parameters are similar. For the HS and HIMS, we fixed the soft photon compactness $l_{s}$ to 1, and for the SIMS, SS and US $l_s$ was set to 10. We fixed the source inclination to 70$^{\circ}$. Starting with a fixed reflection fraction and ionisation parameter ($\xi$) of 0, and an injection index ($\Gamma_{inj}$) equal to 3, we fitted for $l_{h}/l_{s}$ (hard to soft compactness ratio), $\tau_{p}$ (scattering depth) and $l_{nt}/l_{th}$ (fraction of power that goes into accelerating non-thermal particles). Once we obtained parameters consistent with their states, we freed each parameters one by one ($\Gamma_{inj}$, reflection fraction and $\xi$) and re-fit. Depending on the state and flux, we sometimes included an emission line at around 6.4 keV, an edge at around 7.1 keV, and one or two Gaussian absorption components (for the US and SS) to obtain acceptable fits.  We have not placed any constraint on the reflection fraction and allowed it to be greater than 1. As usual, we run the error command for all free parameters. 

\subsection{SMARTS OIR observations and analysis}

Regular optical and near-infrared observations were carried out at Cerro Tololo Inter-American Observatory (CTIO) with the ANDICAM \citep{Depoy03} instrument on SMARTS 1.3m telescope \citep{Subsavage10}. The observations were taken using the Johnson-Kron-Cousin B, V, I optical filters \citep{Bessell98} and the CTIO J and K near-infrared filters \citep{Elias82}. All data reduction and photometry was done with standard IRAF tasks. The optical images went through the process of bias-, overscan-subtraction and flat-field division. In order to have a single exposure infrared image for each filter each night, sky fields were constructed by median combining all images from a given night of the same filter, each dithered image was then sky-subtracted, flat-fielded, aligned to a common reference frame, and finally the processed dithered images were summed. We performed psf photometry on the final images to measure the instrumental magnitude of \SF\ and the comparison stars in the fields. The optical magnitudes of comparison stars were calibrated using the optical primary standard stars \citep{Landolt92} whereas the infrared magnitudes of comparison stars were obtained from The Two Mass All-Sky Survey (2MASS) catalogue \citep{Skrutskie06}. We added 0.05 mag systematic error to all observations.

For calculating the OIR fluxes, we dereddened magnitudes using E(B-V) = 1.3$\pm$0.1 \citep{Orosz97} and extinction laws given by \cite{Cardelli89} with corrections described in \cite{ODonnell94}. The procedure is outlined in detail in \cite{Buxton12}. All measurements used in this work are provided in Table~\ref{table:oir}.

\begin{figure}
\includegraphics[width=88mm]{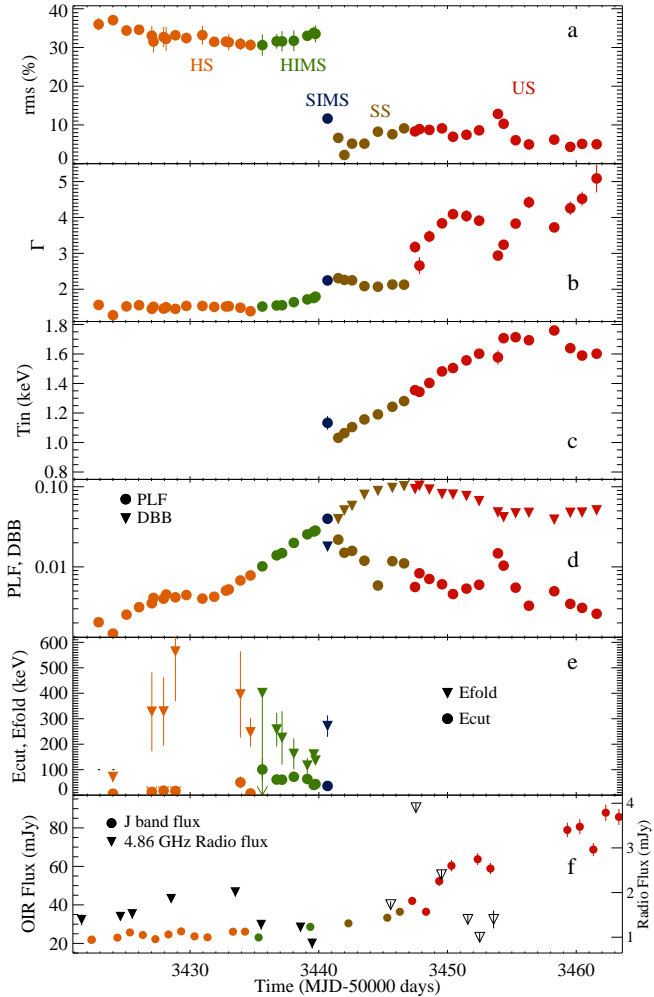}
\caption{
\label{fig:xevol}
The X-ray spectral parameters of the \rxte\ data, fitted with a power-law+diskbb model. a: rms variability, b. X-ray photon index, c: Inner disc temperature, d: Eddington Luminosity fractions of Power-law flux (PLF) and disc blackbody flux (DBB), e: Cut-off and folding energies from \emph{highecut} model, f: $J$ band flux (filled circles), 4.86 GHz radio flux (filled triangles for HS/HIMS, open triangles SS/US). For all figures with multiwavelength evolution, orange, green, blue, brown and red represent the HS, HIMS, SIMS, SS and US, respectively. 
}
%\vspace{-0.85 cm}
\end{figure}

\subsection{VLA Radio observations and analysis}

\SF\ was observed regularly with the VLA throughout the 2005 outburst. The measurements for the HS and HIMS during the earlier parts of the outburst between 2005 February 20 (MJD~53421) to  March 16 (MJD~53445) were published in \citep{Shaposhnikov07}. Here, in addition, we present results of observations over eight epochs from 2005 March 18 (MJD~53447) to April 07 (MJD~53467), as the source was making a transition to the US. During these observations, the array was in the relatively extended B-configuration. The observations were carried out at frequencies of 1.425 and 4.86\,GHz on all epochs, at 8.46\,GHz on all epochs except MJD~53467.4, and one observation at 22.46\,GHz on MJD~53447.5. All observations were taken with 100 MHz of bandwidth, split equally between two 50 MHz channels, and an integration time of 3.3 characterizeds. The data were reduced and imaged following standard reduction procedures within the Common Astronomy Software Application (CASA; \citealt{McMullin07}) software package. 3C286 was used as the primary calibrator, setting the amplitude scale according to the coefficients derived by staff at the National Radio Astronomy Observatory (NRAO). The secondary calibrator J1626$-$2951 was used for the 1.425\,GHz data, while J1607$-$3331 was used for the 4.86 and 8.46\,GHz data, and J1650$-$2943 for the 22.46\,GHz data. To determine the flux density of the source, we fitted the target with a point source in the image plane. The measurements are provided in Table~\ref{table:radio}.

%% /*******************************************************************
%% ** Results                                                        **
%% *******************************************************************/

\section{Results}\label{sec:results}

With the fits to the X-ray data, we describe the X-ray spectral spectral evolution in \S\ref{sub:states}. The disc+power-law model indicated the presence of a high energy excess in some of the US observations. In \S\ref{sub:hard} we investigate the properties of the hard X-ray flares with respect to multiwavelength evolution. We have characterized the X-ray evolution with the \emph{eqpair} model in \S\ref{sub:eqpair} to map the state transitions with changes in the Comptonization properties, test for the presence of non-thermal hybrid plasmas, and determine if the high-energy excess we observed with a power-law fit is due to inadequacies within the phenomenological model. Finally, in \S\ref{sub:SED}, we show the evolution of the radio/OIR flux with respect to the X-ray spectral states and the broad-band SED to understand the relationship between the X-ray, jet, wind and outer disc emission, as well as the emission of the secondary star.

\begin{figure}
\includegraphics[width=88mm]{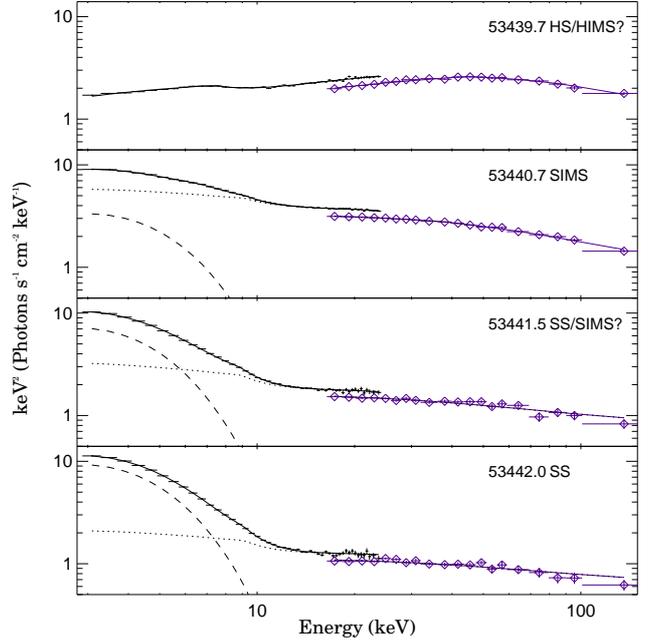}
\caption{\label{fig:sptr1}
The X-ray spectral evolution during the transition from the HS to the SS. Solid lines represent the overall fit, while the dashed line is the disc component and the dotted line is the power-law component.
}
\end{figure}

\begin{figure}
\includegraphics[width=88mm]{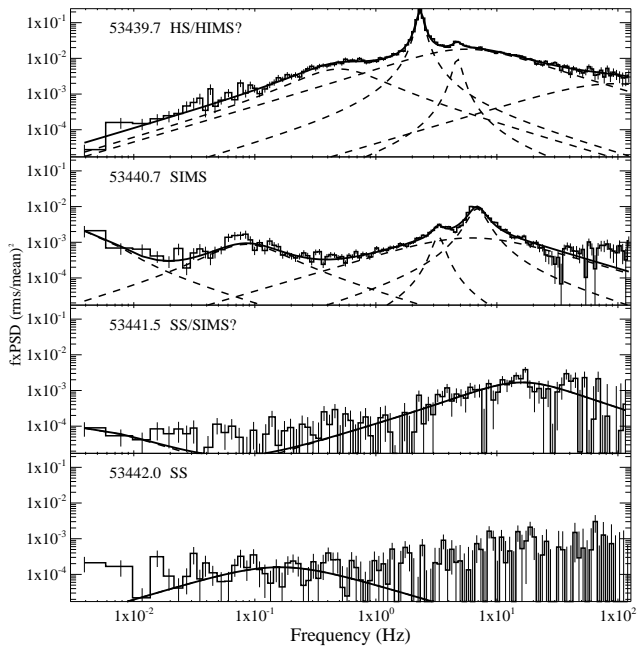}
\caption{\label{fig:ttr1}
The X-ray timing evolution of the power spectral densities (PSD) during the transition from the HS to the SS. The solid line is the overall fit and the dashed lines show the individual Lorentzian components.
}
\end{figure}

\subsection{X-ray spectral evolution and state transitions}
\label{sub:states}

In Fig.~\ref{fig:xevol}, we show the evolution of the key X-ray spectral and temporal parameters that are used to determine the X-ray spectral states. The same plot also shows the evolution of the $J$-band flux, as well as the radio flux at 4.86 GHz. The X-ray spectral fit parameters are given in Table~\ref{table:x1}. 

Some of the spectral/timing state transitions are obvious. The abrupt drop in the rms variability and the appearance of disc emission on MJD~53440.7 marks the transition to the SIMS (shown with blue symbols), while the sudden softening on MJD~53447.5 marks the transition to the US (shown with red symbols). However, the spectral states around the SIMS are more subtle and have been interpreted differently in earlier works. According to the state definitions used in \cite{Motta12} the observations before the SIMS transition are in the HS. However, during this time the X-ray spectrum was softening, and there was a drop in the radio flux. For the last three observations of the HIMS/SIMS (before the transition to the SS), there was an increase in the folding energy of the high-energy cut-off (see Fig.~\ref{fig:xevol}), similar to the case of GX 339-4 \citep{Motta09}. There were also abrupt changes in the Comptonization parameters (see \S~\ref{sub:eqpair}). Therefore, we denote observations between MJD~53435 and MJD~53440 as HIMS (shown with green symbols), as done by \cite{Joinet08} and \cite{Shaposhnikov07}. 

\begin{figure}
\includegraphics[width=88mm]{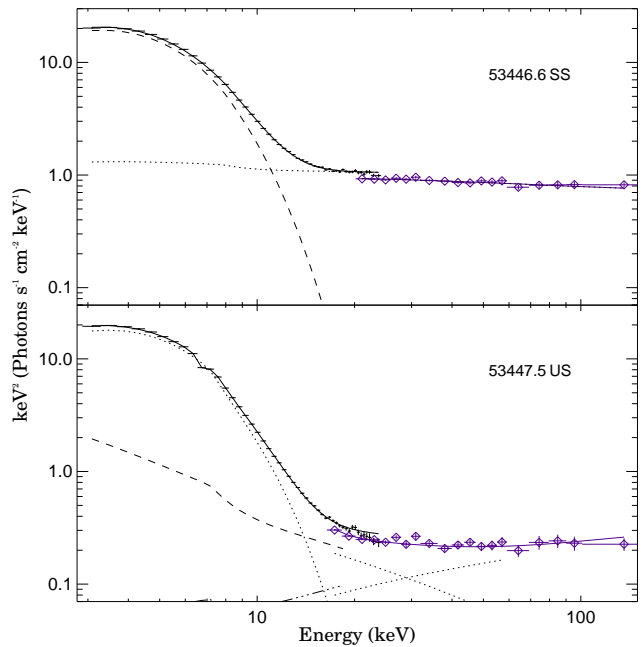}
\caption{\label{fig:sptr2}
The X-ray spectral evolution through the transition from the SS to US. The solid lines are the total fit, while the dashed lines are the disc component and the dotted lines are the power-law components.}
\end{figure}

\begin{figure}
\includegraphics[width=88mm]{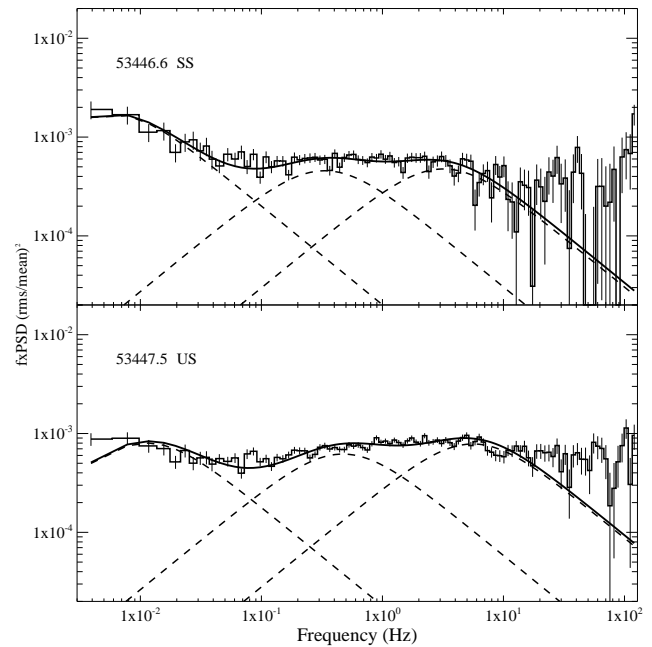}
\caption{\label{fig:ttr2}
X-ray timing evolution of the PSD through the transition from the SS to US. Here, the solid line represents the overall fit and the dashed lines show the individual Lorentzian components.}
\end{figure}

If only the timing characteristics are taken into account, only the observation taken on MJD~53440.7 can be considered to be in the SIMS \citep{Motta12}. In Figs.~\ref{fig:sptr1} and \ref{fig:ttr1} we show the transition from the HS/HIMS to the SS in more detail. In the HIMS, the X-ray spectrum shows the appearance of a disc component (Fig.~\ref{fig:sptr1}) and there is a B-type quasi-periodic oscillation (QPO) in the PSD (Fig.~\ref{fig:ttr1}). In less than a day, the disc then became the dominant component in the soft X-ray band and the B-type QPO disappeared. According to \cite{Motta12}, the observation on MJD~53441.5 was already in the SS (shown by brown symbols; Fig.~\ref{fig:xevol}). However, our results show that it is likely that at this time, the system was in a transition from the SIMS to SS because, even though the B-type QPO had disappeared there was residual broad-band noise and the power-law flux was still significant. We also observed the disappearance of the high energy cut-off after the transition to the SIMS. Finally, the source reached the SS on MJD~53442, broad-band by a much larger disc contribution and smaller power-law contribution. This transition also coincided with the increase in OIR flux. This entire sequence of transitions lasted 2 days.

Figs.~\ref{fig:sptr2} and \ref{fig:ttr2} show the transition from the SS to the US in detail. This transition was marked by a sharp drop in the hard X-ray emission. It is not possible to fit the first (and some of the following) US observations with a single power-law (see \S~\ref{sub:hard}). Interestingly, the timing properties remain similar in 3--30 keV band during the transition. The transition from the SS to the US also coincided with an optically-thin radio flare. 

\subsection{Hard X-ray flares}
\label{sub:hard}

\begin{figure}
\includegraphics[width=88mm]{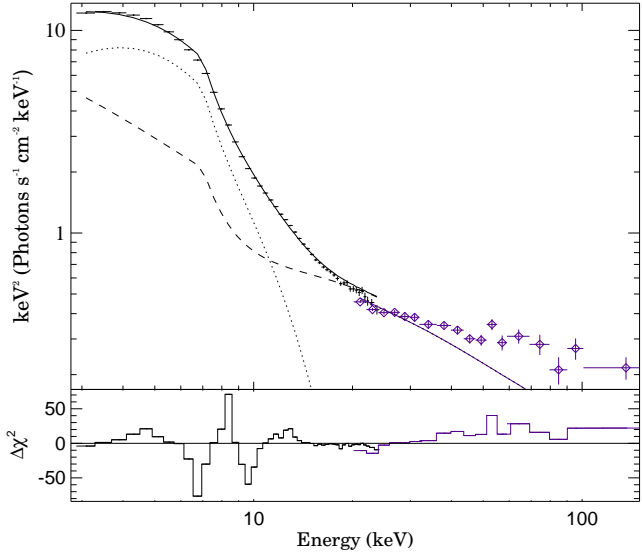}
\caption{\label{fig:resmix}
Top: The unfolded spectrum of one of the US spectra, black crosses represent PCA data and purple diamonds are HEXTE data. We fit the data with an absorbed diskbb (dotted line)+power-law(dashed line) with a smeared edge, where the solid line is the overall fit. Bottom: residuals ($sign[data-model]*\Delta\chi^{2}$) showing that the fit needs a high energy component as well as two absorption lines between 6 and 10 keV.
}
\end{figure}

As discussed in \S~\ref{sub:states}, following the transition to the US, a single power-law component was not adequate to fit the PCA+HEXTE spectrum. Fig.~\ref{fig:resmix} shows the residuals of PCA+HEXTE spectrum when we used a model consisting of a \emph{diskbb}, a single power-law and a smeared edge.  While the PCA data can be fit with a single steep power-law component, above 30 keV a very hard secondary power-law component is required. There were also strong residuals between 6--10 keV which we modelled with two Gaussian components with negative normalizations.

\begin{figure}
\includegraphics[width=88mm]{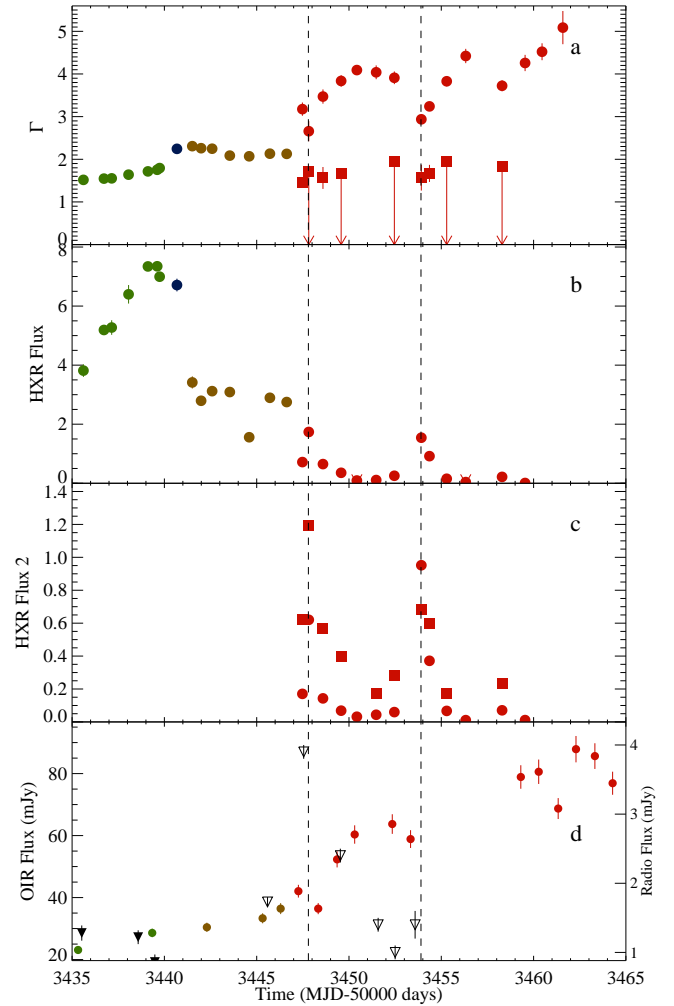}
\caption{\label{fig:hxrevol}
The evolution of the hard X-ray flares. a. The X-ray photon index, where the circles represent the steep power-law component and the squares show the second, hard power-law component. b. The hard X-Ray (25-200 keV) flux in units of $10^{-9}$ ergs cm$^{-2}$ s$^{-1}$. Errors are often smaller than the plot symbol. c. The hard X-ray flux during the US, showing the steep (circles) and hard (squares) power-law fluxes separately. d. The multiwavelength evolution (see Fig.~\ref{fig:xevol} for details). Dashed lines show the times that the hard X-ray flares peaked.
}
\end{figure}

For all US data, we checked for the presence of a secondary power-law component (Fig.~\ref{fig:hxrevol}). In the 15 US observations that we analysed, 10 required an additional hard X-ray component (all before MJD 53460). After this date, the source entered the hyper-soft state where there was no significant hard X-ray emission \citep{Uttley15}. While it is clear from residuals that a second power-law component was required in those 10 observations, due to the degeneracy between model components we were not able to constrain the photon index error range for 5 of these observations and, therefore, only obtained upper limits (Fig.~\ref{fig:hxrevol}a). Following the transition to the US, we observed two hard X-ray flares (see Fig.~\ref{fig:hxrevol}c), one coinciding with the optically-thin radio flare, and the second one coinciding with apparent X-ray spectral hardening (see Fig.~\ref{fig:hxrevol}a). For almost all of our observations, the 25-200 keV flux from the second power-law component dominates the hard X-rays (due to its lower photon index).  

When we compare these hard X-ray flares with the multiwavelength observations we observe that the first X-ray flare is exactly coincident with an optically-thin radio flare on MJD~53447.5. There was also a slight increase in radio emission around MJD~53454. However, the 1.425 GHz and 4.86 GHz radio flux only increased by 1$\sigma$--2$\sigma$. Therefore, we cannot conclusively prove the existence of a second radio flare.

%\vspace{-1.5 cm}

\begin{figure}
\includegraphics[width=88mm]{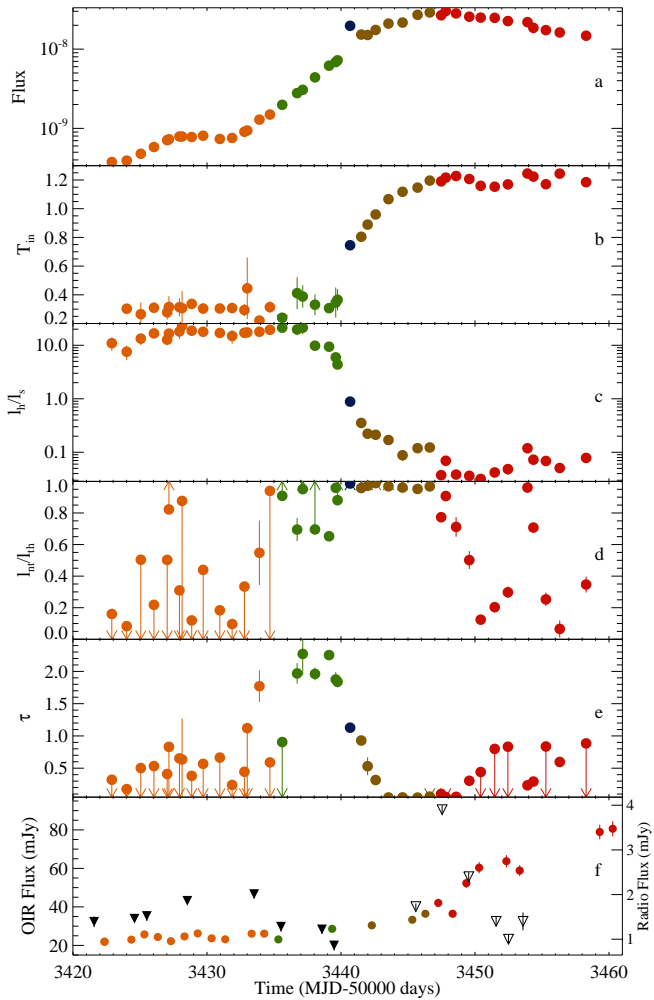}
\caption{\label{fig:eqp1}
The evolution of the key parameters from our \emph{eqpair} fits. a. The unabsorbed 3--25 keV X-ray flux in ergs cm$^{-2}$ s$^{-1}$ b. The inner disc temperature from the \emph{diskpn} model, c. The hard to soft compactness ratio, d. The ratio of non-thermal electrons to thermal electrons, d. The optical depth, and f. The multiwavelength evolution (see Fig.~\ref{fig:xevol} for details).
}
\end{figure}

\begin{figure}
\includegraphics[width=88mm]{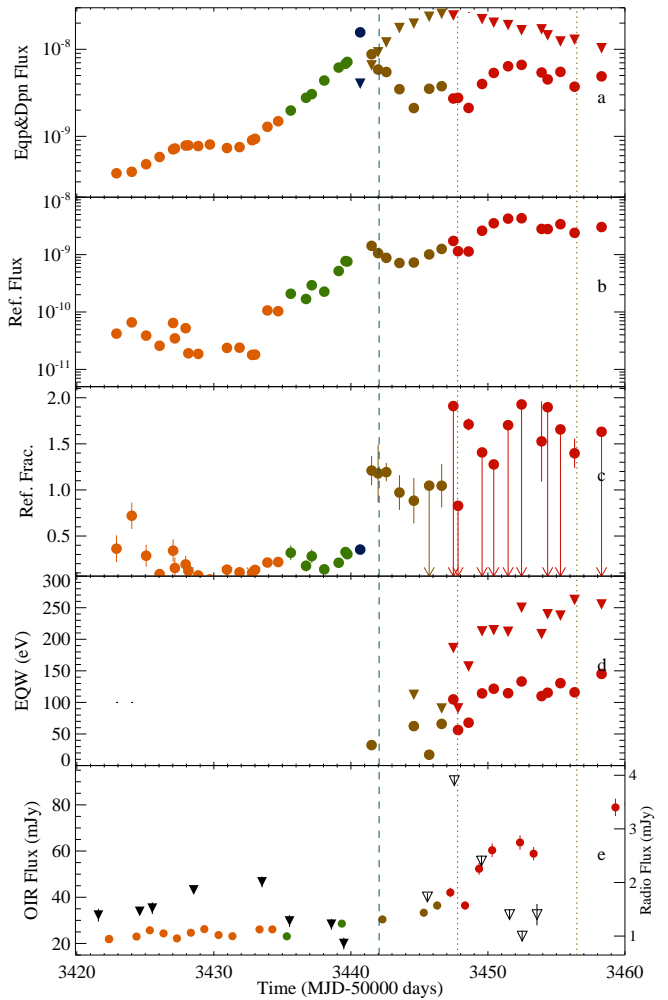}
\caption{\label{fig:eqp2}
The evolution of the key parameters from our \emph{eqpair} fits  a. The 3--25 keV unabsorbed X-ray flux from the \emph{diskpn} (triangles) and \emph{eqpair} (circles) models, where all fluxes are in units of ergs cm$^{-2}$ s$^{-1}$, b. The flux from the reflection in the \emph{eqpair} model (circles). c. The reflection fraction, d. Equivalent widths of the \wsim6.7 keV (circles) and \wsim8 keV (triangles) absorption lines. e. Multiwavelength evolution. Dashed line indicate the time of the \chandra\ observation, while dotted lines indicate the times of the \xmm\ observations.
}
\end{figure}

\subsection{Comptonization fits, case for hybrid corona}
\label{sub:eqpair}

While one possibility is that there were two separate power-law components in the US originating from different sources, it is also possible that a single power-law simply does not provide an adequate description of a Comptonizing corona in the case of \SF. We may not have two separate components, but may just require a comprehensive Comptonization model that includes reflection. Therefore, we fit all of our spectra with \emph{eqpair}, because it is the most comprehensive Comptonization model that allows a hybrid plasma containing a combination of thermal and non-thermal electron energy distributions. The evolution of important parameters are shown in Figs.~\ref{fig:eqp1} and \ref{fig:eqp2}, along with evolution of the multiwavelength parameters. Key parameters are provided in Table~\ref{table:eqp}.

As mentioned earlier (\S\ref{sub:hard}), as well as in other works \citep{Uttley15, Shidatsu16}, possibly due to complex wind structures in the 6--10 keV region (during the SS and US), our model required one or two Gaussians between 6.6--7.2 keV and 7.6--8.2 keV, as well as an edge between 8.5--9.5 keV (based on \citealt{DiazTrigo07}) to obtain reasonable fits. The resolution of the PCA does not allow for a detailed study of the energies and equivalent widths (EW) of the absorption features. On the other hand, during the ``Obs 1" of \cite{DiazTrigo07} as indicated by the dotted line on MJD 53448 in Fig.~\ref{fig:eqp2}, our measured EWs are similar to those obtained by \xmm\ (see Table~\ref{table:eqp}). However, this was not the case for ``Obs 2" where we observe much larger EWs. Therefore, we can only claim that these features are required in the fit, but the measurements of EWs may be incorrect. Assuming general trends are valid, the data indicate that these lines became more prominent as the source approached the hyper-soft state.

The general trend in the \emph{eqpair} parameter evolution is consistent with results from other GBHTs \citep{delSanto08, delSanto16}, where the hard state had a low disc temperature, a high hard-to-soft compactness ratio ($l_{h}/l_{s}$), low optical-depth, and almost all observations required a thermal energy distribution with $l_{nt}/l_{th}$ close to zero (except one which may be finding a local minima, possibly due to degeneracy of the fit parameters), as well as a small reflection fraction. In the HIMS, the disc temperature, optical depth and reflection fraction increased, while the $l_{h}/l_{s}$ decreased, which are all consistent with the inner disc approaching the inner-most stable orbit around the black hole. The fits required a hybrid model to completely account for the non-thermal electron energy distribution, which is common for intermediate states \citep{delSanto08, Gierlinski99, Malzac06}. The SIMS and SS required much larger inner disc temperatures and $l_{h}/l_{s}$ decreased further before it plateaued. The optical depth also decreased and the reflection fraction was close to (and sometimes slightly higher than) 1. During these states, the fits indicate a corona dominated by non-thermal electron energy distribution.

\begin{figure}
\includegraphics[width=88mm]{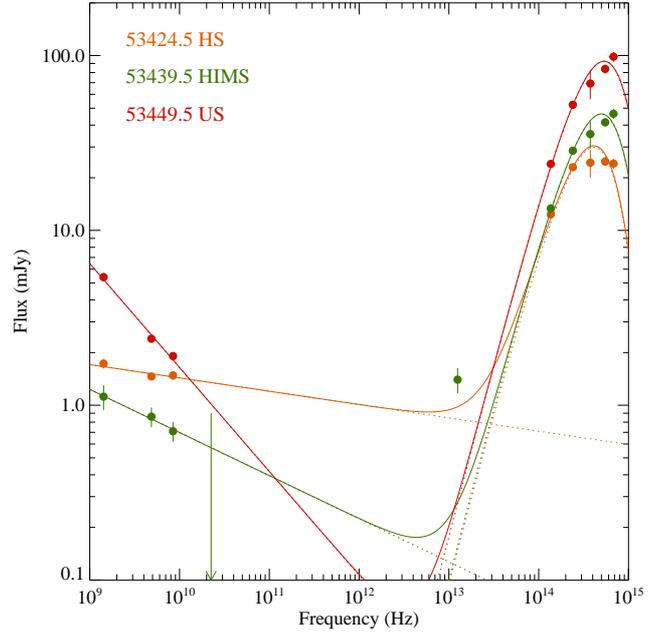}
\caption{\label{fig:seds}
The SEDs of chosen observations in HS, HIMS and US. Same colouring is used as in previous plots with orange representing HS, green HIMS and red US. The model is a power-law (representing jet emission) + blackbody (representing roughly the outer disc + companion star). The 24$\mu$m ($1.25 \times 10^{13} Hz$) \emph{Spitzer} observation on MJD~53439.5 and all upper limits are not included in the fit procedure.
}
\end{figure}

\begin{figure}
\includegraphics[width=88mm]{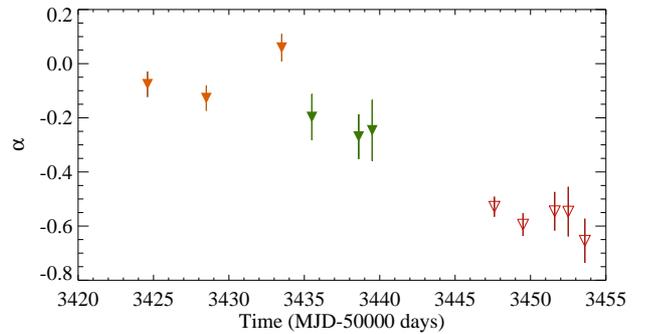}
\caption{\label{fig:radind}
Evolution of the radio spectral index $\alpha$ at different spectral states. Orange filled triangles in the HS, green filled triangles in the HIMS and red open triangles in the US. 
}
\end{figure}

%\begin{figure}
%\includegraphics[width=88mm]{f10.eps}
%\caption{\label{fig:sedhshims}
%The SED of the observations in the HS and HIMS. The model is a power-law (representing jet emission) + blackbody (representing roughly the outer disk + companion star). For the observation on MJD~53433.5, the power-law index is fixed to 0, whereas the index is kept free for all other observations. The 24$\mu$m \emph{Spitzer} observation on MJD~53439.5 and all upper limits are not included in the fit procedure. The diamonds show the quiescence flux levels from Greene et al. 2001.
%}
%\end{figure}

%\begin{figure}
%\includegraphics[width=88mm]{f11.eps}
%\caption{\label{fig:sedssus}
%The SED of observations in the SS (MJD~53445.5) and the US. The model is a power-law (representing jet emission) + blackbody (representing roughly the outer disk + companion star). The diamonds show quiescence flux levels from Greene et al. 2001.
%}
%\end{figure}

Importantly, our results show that when a high reflection fraction is allowed (as seen from the trend in Fig.~\ref{fig:eqp2}), the fits do not require a secondary high energy component in the US (the reduced $\chi^{2}$ values remain below 2). 

Secondly, the observations show that for the two hard X-ray flares observed around MJDs~53449 and 53454, a non-thermal electron energy distribution was necessary. This is more clear during the second X-ray flare, close to MJD~53454, where $l_{nt}/l_{th}$ increased sharply, along with a jump in $l_{h}/l_{s}$, as the inner disc temperature also increased. 

\subsection{SED evolution}
\label{sub:SED}

The excellent optical NIR coverage and reasonably well sampled radio coverage allowed us to investigate the evolution of the spectral energy distribution (SED) during this outburst and relate the changes to the X-ray spectral states and X-ray evolution. Since detailed SED studies with realistic jet models \citep{Migliari07} and irradiated disc models \citep{Shidatsu16} are presented elsewhere, here we will concentrate on the evolution of radio spectrum.

Fig.~\ref{fig:seds} shows the broad-band spectra during three representative epochs (one in each of the HS, HIMS and US). The data was modelled with a power-law (representing jet emission) + blackbody (representing roughly the outer disc). The evolution of radio power-law spectral indices ($\alpha$) are shown in Fig.~\ref{fig:radind}. We used 1.425 Hz, 4.86 and 8.46 Hz fluxes for the fits. Due to low source elevation and poor weather, we find significant phase decorrelation at high radio frequencies (22.46 GHz) on MJDs~53428.5, 53438.5 and 53439.5. By treating every other scan of the phase calibrator as the target, we estimate that the phase calibrator observations (at 22.46 GHz) suffer \wsim\ 20\% phase decorrelation. However, \SF\ is \wsim 10$^{\circ}$ lower in elevation than the phase calibrator meaning they would be further affected. Therefore, we do not use these upper limits in our fits. We included \emph{Spitzer} MIPS data at 24 $\mu$m on MJD~53439.5 \citep{Migliari07} in the Fig.~\ref{fig:seds}, but not in the fit. 

The radio spectrum was close to flat ($\alpha \approx 0$) in the HS, before steepening in the HIMS. In the US, during the bright radio flare, the radio spectral index steepened significantly, indicating an optically-thin jet spectrum (see Fig.~\ref{fig:radind}). In the near-infrared to optical, the flux in all bands increased as the source evolved from the HS towards the US (see Fig.~\ref{fig:seds}).

%% XX we have to define radio spectral index in the introduction
%We find that MJD~53447.5 observations were not significantly affected.

%In Figs.~\ref{fig:sedhshims} and \ref{fig:sedssus}, we show the mean emission level measured in quiescence, when the stellar flux dominates. The quiescent emission from the companion is a significant fraction of total OIR emission in the HS. For the observation on MJD~53433.5 the power-law index was fixed to 0, otherwise the fit tends to connect to OIR points with an unrealistic index. We included \emph{Spitzer} MIPS data at 24 $\mu$m on MJD 53439.5 \citep{Migliari07} in the figure, but not in the fit. 

%During the SS and US, the radio flare is optically-thin (Fig.~\ref{fig:seds} and Table~\ref{table:radio}) and shows similar characteristics of the bright, fast, jets observed during the state transitions of many GBHTs. The blackbody component roughly fits the X-ray data, and its strength increases significantly as the source approaches the hypersoft state. 

%% /*******************************************************************
%% ** Discussion                                                                      **
%% *******************************************************************/

\section{Discussion}\label{sec:discussion}

\subsection{State transitions}

While the general spectral and timing evolution, as well as the spectral state identification of this source have previously been discussed \citep{Shaposhnikov07, Motta12}, here, we discuss the evolution in terms of the \emph{eqpair} parameters and their relation to the multiwavelength evolution. First of all, the transition out of the hard state was marked by an increase in the optical depth ($\tau$) and an increase in the non-thermal electron distribution ($l_{nth}/l_{th}$). During the transition, the radio flux decreased and the radio spectrum became optically-thin as shown in Figs.~\ref{fig:seds} and \ref{fig:radind} (also see \citealt{Shaposhnikov07}). A similar evolution was observed in H1743$-$322 \citep{MillerJ12}, which showed radio spectral index softening and a slight drop in radio flux during the HIMS before the quenching of the compact jet and the launching of an optically-thin radio flare. While the H1743$-$322 radio flare peaked during the SS, VLBA observations indicated that the time of launch was close to the transition from the HIMS to the SIMS, which was a few days before the peak radio flux. It was not possible to do a similar analysis for \SF\, because the jet was not resolved. However, if we assume similar time scales it is possible that the optically-thin ejecta were launched earlier in this system. Such a delay would make the radio and hard X-ray flares out of synch, with the radio preceding the hard X-ray flare. %Such a case will be difficult to understand if all jets require a geometrically thick medium to be launched.

\cite{Corbel12} showed that the radio flux and radio spectral index gradually increasing (becoming flat and then inverted) as the compact jets were re-launched during the outburst decay of GX~339$-$4. MAXI~J1836$-$194 showed a similar evolution, where the radio spectrum softened as the source entered the HIMS from the HS, and then became highly inverted again in the hard state during the decay \citep{Russell13a, RussellT14}. A natural interpretation of this would be that the jets become more collimated and compact during the transition to the hard state during the decay, possibly as the magnetic flux accumulates close to the inner parts of the accretion flow. H1743$-$322 and this source show that perhaps the reverse evolution is taking place during the rise, that the magnetic flux diffuses out faster than it can be accumulated \citep{Begelman14}, reducing the power and collimation of the jet. However, the relatively strong flux at 24 $\mu$m during the HIMS (on MJD~53439.5) might suggest that this was not occurring in \SF. As discussed in \cite{Migliari07} and shown in Fig.~\ref{fig:seds}, it is difficult to explain the flux level as emission from the outer parts of the accretion disc or as dust from a circumbinary disc because the emission was variable and much stronger than what has been observed in other sources \citep{Muno06}. If it was coming from the jet, the radio spectrum cannot be fit with a single power-law, and may include multiple components as the compact jet was quenching.

%%XX throw away sentence, see TDR, does not say anything new

As the source made its transition to the SS, the non-thermal compactness ratio peaked at a level of 1 and remained steady while the optical depth decreases. As expected, the hard-to-soft compactness ratio decreased as well. At this time, the reflection fraction was $\sim$1, indicating a compact corona and an inner disc that was close to the black hole. In our \emph{power-law}+\emph{diskbb} fits, this transition showed an increase in the folding energy of the \emph{highecut} component, indicating higher and higher cut-off energies \citep{Joinet08} as the source moved towards the SS, which is in agreement with the increasing $l_{nth}/l_{th}$. Such behaviour has been observed during the state transitions of GX~339$-$4 and Swift~J1745$-$26 \citep{delSanto16} and can be explained by the presence of a dead-zone in the intermediate states in the elevated disc model of \cite{Begelman15}.

The SS to US transition was coincident with an optically-thin radio flare (though the actual ejection may have preceded the transition) and an increase in the OIR flux. As the source evolved in the SS, the $l_{nth}/l_{th}$ decreased and the electrons thermalize. However, during the hard X-ray flare, the non-thermal compactness ratio increased up to unity, with a slight increase in hard-to-soft compactness ratio. Along with high reflection fraction, a single hybrid Comptonization component was adequate to represent the X-ray spectrum. We note that the reflection fraction was not well constrained because the lower energies of the reflection component is in the part of the spectrum with the iron absorption lines and edges (where the \rxte\ data has a higher effective area), and the resolution of \rxte\ makes it impossible to resolve each component. Nevertheless, it is not clear how the electrons became non-thermal and then thermal again in the US on those time scales.  A possible explanation is the disc breaking scenario of \cite{Nixon14}. Since the inclination and spin angles are misaligned in \SF, the mechanism described in \cite{Nixon14} may be able to heat up the disc over the time scales observed here.

\subsection{Hard X-ray flares and radio emission}

We have identified two hard X-ray flares (see Fig.~\ref{fig:hxrevol}) during the rise of the 2005 outburst of \SF. The first flare occurred during the transition from the SS to the US. This transition also coincided with an optically-thin radio flare similar to the transient jets observed in many GBHTs \citep{Dhawan00, Fender06, Gallo10}. The association of the second hard X-ray flare with a radio flare is not as clear due to lack of radio observations between MJD~53455 and MJD~53460. Historically, \SF\ has shown many instances of luminous radio flares (relativistic ejections) coinciding with hard X-ray flares. But the association cannot be examined in detail in these older data due to limited spectral capability of the BATSE instrument on \emph{CGRO}. For example, between MJD~49550 and MJD~49700, three radio flares were observed to be coincident with X-ray flares (where the X-ray peaked earlier than the radio, \citealt{Harmon95}), but no radio flare was observed within the next year even though several hard X-ray flares took place \citep{Tavani96}. However, it is possible that radio ejection events did occur between MJD~49700 and MJD~50000 but were simply missed due to the timing of the radio observations and because the relation between spectral states and radio jets was not well known at the time and it was not easy to determine the X-ray spectral state. 

Aside from the historical note, the important observation here is the clear association of the radio jet with the hard X-rays. This association is obvious in the case of compact jets, which are always associated with the hard X-ray spectral state, and can only turn back on (following their quenching in the soft state) when the X-ray spectrum has hardened sufficiently during the outburst decay \citep{Kalemci13}. However, this association is less clear for the optically-thin flares during the outburst rise. \cite{Fender09} investigated the relationship between the spectral hardness and major radio ejections and found that while the association is complex, at least in XTE~J1859+226, a fast hardening is associated with a major flare event. This association can be related to the production, transport and dissipation of magnetic fields in the inner disc. The presence of hard X-rays emission indicates the presence of a geometrically-thick corona, which makes it easier to produce \citep{Begelman14, Kylafis15} and transport \citep{Beckwith09} magnetic flux. Similarly, with the compact jets, the presence of some form of hot, vertically-extended accretion flow may be a necessary condition for an optically-thin radio flares as well.

\subsection{OIR evolution, radio flare and winds}
\label{sub:disOIR}

Explaining the behaviour of OIR emission from \SF\ is a difficult task due to sub-giant secondary contributing significant emission, especially in the HS and HIMS. For other well studied systems, the OIR emission is dominated by the compact jet during the hard state rise and decay (e.g. \citealt{Kalemci13}, but also see \citealt{Veledina13} for an alternative explanation based on a hot-flow model). In fact, for GX~339$-$4 in several outbursts \citep{Coriat09}, MAXI~J1836$-$194 \citep{Russell13a, RussellT14}, and XTE~J1550$-$564 (Kalemci et al. in preparation), the OIR emission drops down significantly as the source enters the HIMS. On the other hand, we observe no decrease in the OIR flux for \SF\ during the HIMS (although it is clear that the jet flux is decreasing, and perhaps becoming optically-thin at this time as shown in Fig.~\ref{fig:seds}), in fact it rises as the source enters the SS and then the US. This peculiar behaviour has also been discussed by \cite{Shidatsu16}. disc size cannot explain this difference as the binary separations of \SF, GX 339$-$4 and XTE~J1550$-$564 are similar. The only difference between them is the high inclination of \SF\, whereas the other sources are low inclination. A possible explanation is provided in \cite{Shidatsu16}, with the scattering in a strong wind increasing the irradiation and making the disc brighter. 

With the PCA observations, we infer the presence of winds from the beginning of the SS and beyond based on the detection of absorption lines. Our first detection is on MJD~53441.5, around the same time as the \chandra\ observation (Obsid 5460), which started on MJD~53441.9 \citep{Neilsen12}. The date of the \chandra\ observation is shown in Fig.~\ref{fig:eqp2} with a dashed line. In \cite{Neilsen12} this observation is described to be moving out of the hard state, while in \cite{Neilsen13}, it is simply described as an observation in the hard state. Our analysis, as well as earlier timing and spectral analysis, indicate that at this time, the source had already left the hard state and was completely in the soft state by MJD~53442.0 (Figs.~\ref{fig:sptr1} and \ref{fig:ttr1}). This observation is ``harder" than the other \chandra\ observation on MJD~53461.5, which was taken in the extremely soft hypersoft state.

%% XX need to talk about homan?
An interesting fact overlooked by earlier works is that the optically-thin radio flare, which peaked at around MJD~53447, coexisted with the disc wind detected by both \chandra\ \citep{Neilsen13} and \xmm\ \citep{DiazTrigo07}. The times of \chandra\ and \xmm\ observations are indicated by dashed and dotted lines, respectively, in Fig.~\ref{fig:eqp2}. While the compact jet / wind dichotomy is well documented \citep{Ponti12}, this is one of the rare cases that a wind and a jet of some form are observed together in a GBHT \citep[a recent case is the discovery of deep $H$ and $He$ $P-Cyg$ profiles existing along with radio emission from a compact jet in V404 Cyg,][]{MunozDarias16}. In the $\beta$ state of GRS~1915+105 (which generally show a soft X-ray spectrum) strong winds are observed, whereas no winds are observed in the hard states. Based on this, it was claimed that the intense mass loss due to winds were prohibiting the launching of the jets in this source by halting flow of matter into the compact jet \citep{Neilsen09}. Further analysis indicated that the winds were quenched during the dips (when the jets are presumably launched, \citealt{Mirabel98}), but were strong and fast in the flaring part of the $\beta$-state \citep{Neilsen12b}. Given that the winds are launched tens of thousands of gravitational radii from the black hole, it is more natural to assume that changes in inner accretion flow regulate the outflows, and it is not surprising to observe jets and winds together in transitional states. We note that \SF\ showed optically-thin radio flares in earlier outbursts for which the flux densities reach as high as 10 Jy, and were usually larger than 100 mJy, at 1.49 Ghz  \citep{Hjellming95, Harmon95}. The 2005 outburst on the other hand only reached \wsim 6 mJy at its peak. Because the radio coverage was almost daily, it is unlikely that an order of magnitude larger radio peak was missed. Therefore, in the case of \SF, a weak wind was observed together with a weak optically-thin radio jet tapping the same accretion power reservoir.

%\section{Summary and Conclusions}

%% /*******************************************************************
%% ** Acknowledgments                                               **
%% *******************************************************************/

\section*{Acknowledgements}
 E.K acknowledges T\"UB\.ITAK BIDEB 2219 Award that enabled a sabbatical visit at JILA, CU Boulder and is grateful to Mitch Begelman, Greg Salvesen, Chris Nixon and Phil Armitage for their hospitality and extended discussions on state transitions. A significant portion of this work was completed at JILA, using the computer infrastructure of CU Boulder during EK's sabbatical visit. EK also acknowledges support from T\"UB\.ITAK 1001 Project 115F488. TDR thanks James Miller-Jones and Peter Curran on the interpretation of the radio data.
 
%% /*******************************************************************
%% ** Bibliography                                                   **
%% *******************************************************************/

%\bibliographystyle{mn2e_fixed}
%\bibliography{./refs}

%%\clearpage
%%%%%%%%%%%%%%%%%%%%%%%%%%%%%%%%%%%%%%%%%%%%%%%%%%

%%%%%%%%%%%%%%%%% APPENDICES %%%%%%%%%%%%%%%%%%%%%

%\clearpage

\appendix

%\makesavenoteenv{tabular}
%\makesavenoteenv{table}

\begin{table*}
  \centering 
  \caption{List of observations IDs, observation times, spectral states and \emph{Power-law + diskbb} spectral fit parameters}
  \label{table:x1}
  \begin{tabular}{lcccccccccc}
   \hline
\# &Obsid &Date &RMS &$\Gamma^{a}$ &$T_{in}$ &PLF$^{b}$ &DBB$^{c}$ &$E_{Cut}$ &$E_{fold}$ & State \\
 &  &MJD (days) &\% &  &(keV) &  &  &(keV) &(keV) & \\ \hline
1 &90058-16-02-00 & 53422.9 & 35.99$\pm$1.61 & 1.56$\pm$0.03 & - &  0.37 &  - & - & - & HS \\
2 &90058-16-03-00 & 53424.0 & 37.07$\pm$1.40 & 1.28$\pm$0.05 & - &  0.40 &  - &   5.5$\pm$  1.4 &  71.2$\pm$ 19.9 & HS\\
3 &90058-16-04-00 & 53425.1 & 34.38$\pm$1.46 & 1.52$\pm$0.02 & - &  0.48 &  - & - & -& HS \\
4 &90428-01-01-00 & 53426.0 & 34.59$\pm$0.78 & 1.55$\pm$0.01 & - &  0.58 &  - & - & - & HS\\
5 &90058-16-05-00 & 53427.0 & 33.05$\pm$2.56 & 1.45$\pm$0.03 & - &  0.72 &  - & $<$ 12.5 & 327.4$\pm$154.7 & HS\\
6 &90428-01-01-01 & 53427.2 & 31.53$\pm$2.82 & 1.51$\pm$0.02 & - &  0.73 &  - & - & - & HS\\
7 &90058-16-07-00 & 53427.9 & 32.70$\pm$2.60 & 1.46$\pm$0.03 & - &  0.80 &  - & $<$ 17.7 & 328.4$\pm$134.0 & HS\\
8 &90428-01-01-03 & 53428.1 & 32.22$\pm$3.12 & 1.50$\pm$0.02 & - &  0.79 &  - & - & - & HS\\
9 &90428-01-01-04 & 53428.9 & 33.15$\pm$0.94 & 1.45$\pm$0.01 & - &  0.78 &  - & $<$ 16.3 & 563.5$\pm$194.4 & HS\\
10 &90428-01-01-02 & 53429.7 & 32.44$\pm$1.38 & 1.53$\pm$0.01 & - &  0.80 &  - & - & - & HS\\
11 &90428-01-01-05 & 53431.0 & 33.19$\pm$2.49 & 1.53$\pm$0.01 & - &  0.73 &  - & - & - & HS \\
12 &90428-01-01-09 & 53431.9 & 31.47$\pm$0.86 & 1.51$\pm$0.01 & - &  0.75 &  - & - & - & HS\\
13 &90428-01-01-10 & 53432.8 & 31.51$\pm$1.27 & 1.52$\pm$0.01 & - &  0.91 &  - & - & - & HS\\
14 &91404-01-01-00 & 53433.0 & 31.36$\pm$2.13 & 1.53$\pm$0.01 & - &  0.94 &  - & - & - & HS\\
15 &91404-01-01-02 & 53433.9 & 30.92$\pm$1.59 & 1.48$\pm$0.01 & - &  1.29 &  - &  50.2$\pm$ 23.0 & 395.3$\pm$169.4 & HS\\
16 &91404-01-01-03 & 53434.7 & 30.64$\pm$1.44 & 1.39$\pm$0.02 & - &  1.51 &  - &   6.6$\pm$  2.4 & 246.5$\pm$ 56.1 & HS\\
17 &91404-01-01-01 & 53435.6 & 30.62$\pm$2.76 & 1.52$\pm$0.01 & - &  1.98 &  - & 100.7$\pm$ 43.4 & $<$400.6 & HIMS\\
18 &91702-01-01-00 & 53436.7 & 31.59$\pm$1.91 & 1.55$\pm$0.01 & - &  2.79 &  - &  61.3$\pm$ 10.7 & 257.6$\pm$ 66.2 & HIMS\\
19 &91702-01-01-02 & 53437.1 & 31.59$\pm$2.62 & 1.55$\pm$0.02 & - &  3.07 &  - &  60.7$\pm$ 20.2 & 224.8$\pm$104.4 & HIMS\\
20 &91702-01-01-03 & 53438.1 & 31.74$\pm$2.68 & 1.64$\pm$0.01 & - &  4.40 &  - &  72.1$\pm$ 14.1 & 161.9$\pm$ 60.9 & HIMS\\
21 &91702-01-01-05 & 53439.1 & 33.03$\pm$0.70 & 1.72$\pm$0.01 & - &  6.17 &  - &  63.6$\pm$ 10.3 & 115.9$\pm$ 30.5 & HIMS\\
22 &90704.04-01-01 & 53439.6 & 33.79$\pm$1.61 & 1.75$\pm$0.01 & - &  6.88 &  - &  40.0$\pm$  5.2 & 159.4$\pm$ 18.4 & HIMS\\
23 &90704-04-01-00 & 53439.7 & 33.52$\pm$2.21 & 1.79$\pm$0.01 & - &  7.25 &  - &  43.4$\pm$  3.5 & 135.7$\pm$ 13.0 & HIMS\\
24 &91702-01-02-00G & 53440.7 & 11.64$\pm$0.88 & 2.24$\pm$0.02 & 1.13$\pm$0.05 & 16.08 &  3.69 &  36.1$\pm$  6.5 & 271.1$\pm$ 42.3 & SIMS\\
25 &91702-01-02-01 & 53441.5 &  6.65$\pm$1.18 & 2.31$\pm$0.03 & 1.03$\pm$0.01 &  8.32 &  6.96 & - & - & SS\\
26 &91702-01-02-03 & 53442.0 &  2.23$\pm$0.16 & 2.26$\pm$0.03 & 1.06$\pm$0.01 &  5.66 &  9.44 & - & - & SS\\
27 &91702-01-02-06 & 53442.6 &  5.18$\pm$0.40 & 2.25$\pm$0.03 & 1.10$\pm$0.01 &  6.03 & 11.44 & - & - & SS\\
28 &91702-01-03-00 & 53443.5 &  5.17$\pm$0.21 & 2.09$\pm$0.02 & 1.16$\pm$0.01 &  4.13 & 16.88 & - & - & SS\\
29 &91702-01-04-00 & 53444.6 &  8.25$\pm$1.09 & 2.07$\pm$0.03 & 1.19$\pm$0.01 &  2.03 & 19.62 & - & - & SS\\
30 &91702-01-05-00 & 53445.7 &  7.59$\pm$0.31 & 2.13$\pm$0.02 & 1.24$\pm$0.01 &  4.33 & 22.78 & - & - & SS\\
31 &91702-01-05-01 & 53446.6 &  9.12$\pm$0.50 & 2.13$\pm$0.03 & 1.28$\pm$0.01 &  4.10 & 24.94 & - & - & SS\\
32 &91702-01-06-00 & 53447.5 &  8.32$\pm$0.17 & 3.18$\pm$0.16 & 1.35$\pm$0.01 &  2.50 & 24.20 & - & - & US\\
33 &91702-01-06-01 & 53447.8 &  8.93$\pm$0.53 & 2.66$\pm$0.24 & 1.34$\pm$0.01 &  3.27 & 26.48 & - & - & US\\
34 &91702-01-07-00 & 53448.6 &  8.72$\pm$0.17 & 3.47$\pm$0.17 & 1.40$\pm$0.01 &  3.50 & 24.73 & - & - & US\\
35 &91702-01-08-00 & 53449.6 &  9.11$\pm$0.25 & 3.84$\pm$0.14 & 1.48$\pm$0.02 &  3.40 & 22.53 & - & - & US\\
36 &91702-01-09-00 & 53450.4 &  6.94$\pm$0.13 & 4.09$\pm$0.10 & 1.50$\pm$0.02 &  2.95 & 22.42 & - & - & US\\
37 &91702-01-10-00 & 53451.5 &  7.45$\pm$0.17 & 4.04$\pm$0.16 & 1.56$\pm$0.02 &  3.27 & 21.89 & - & - & US\\
38 &91702-01-11-00 & 53452.5 &  8.60$\pm$0.26 & 3.91$\pm$0.15 & 1.60$\pm$0.02 &  3.55 & 19.36 & - & -& US \\
39 &91702-01-12-00 & 53453.9 & 12.84$\pm$0.55 & 2.94$\pm$0.07 & 1.58$\pm$0.05 &  7.97 & 13.80 & - & - & US\\
40 &91702-01-13-00 & 53454.4 & 10.30$\pm$0.43 & 3.24$\pm$0.07 & 1.71$\pm$0.03 &  5.77 & 12.74 & - & - & US\\
41 &91702-01-14-00 & 53455.3 &  6.04$\pm$0.27 & 3.83$\pm$0.12 & 1.71$\pm$0.02 &  3.34 & 14.29 & - & - & US\\
42 &91702-01-15-00 & 53456.3 &  4.96$\pm$0.53 & 4.42$\pm$0.17 & 1.69$\pm$0.02 &  2.12 & 14.39 & - & - & US\\
43 &91702-01-16-00 & 53458.3 &  6.20$\pm$0.21 & 3.72$\pm$0.13 & 1.76$\pm$0.03 &  2.93 & 12.07 & - & - & US\\
44 &91702-01-17-00 & 53459.5 &  4.34$\pm$0.96 & 4.26$\pm$0.19 & 1.64$\pm$0.03 &  2.23 & 14.08 & - & - & US\\
45 &91702-01-18-01 & 53460.5 &  5.11$\pm$0.52 & 4.52$\pm$0.20 & 1.59$\pm$0.02 &  1.99 & 13.82 & - & - & US\\
46 &91702-01-19-00 & 53461.6 &  5.01$\pm$0.23 & 5.09$\pm$0.39 & 1.60$\pm$0.02 &  1.69 & 14.89 & - & - & US\\
\hline
\multicolumn{11}{l}{$^{a}$ Photon index}\\
\multicolumn{11}{l}{$^{b}$ Flux from \emph{power-law} component in 3-25 keV band in units of $10^{-9}$ ergs cm$^{-2}$ s$^{-1}$}\\
\multicolumn{11}{l}{$^{c}$ Flux from \emph{diskbb} component in 3-25 keV band in units of $10^{-9}$ ergs cm$^{-2}$ s$^{-1}$}\\
\end{tabular}
\end{table*}

\begin{table*}
  \centering 
  \caption{SMARTS Optical and Near Infrared Measurements}
  \label{table:oir}
  \begin{tabular}{lccccc}
   \hline
Date &B mag &V mag &I mag &J mag &K mag \\
MJD (day) &  &  &  &  &  \\ \hline
53422.3 & 18.530$\pm$0.061 & 17.017$\pm$0.063 & 15.019$\pm$0.066 & 13.345$\pm$0.055 & 12.349$\pm$0.057 \\
53424.4 & 18.451$\pm$0.062 & 16.947$\pm$0.061 & 14.948$\pm$0.065 & 13.290$\pm$0.056 & 12.292$\pm$0.056 \\
53425.3 & 18.217$\pm$0.065 & 16.758$\pm$0.057 & 14.784$\pm$0.058 & 13.169$\pm$0.055 & 12.182$\pm$0.055 \\
53426.3 & 18.268$\pm$0.062 & 16.794$\pm$0.062 & 14.849$\pm$0.072 & 13.227$\pm$0.056 & 12.199$\pm$0.059 \\
53427.3 & 18.595$\pm$0.068 & 17.055$\pm$0.061 & 15.010$\pm$0.067 & 13.328$\pm$0.060 & 12.297$\pm$0.061 \\
53428.3 & 18.214$\pm$0.065 & 16.786$\pm$0.060 & 14.838$\pm$0.062 & 13.214$\pm$0.056 & 12.244$\pm$0.059 \\
53429.3 & 18.182$\pm$0.077 & 16.724$\pm$0.061 & 14.765$\pm$0.059 & 13.147$\pm$0.057 & 12.171$\pm$0.057 \\
53430.3 & 18.411$\pm$0.063 & 16.908$\pm$0.061 & 14.893$\pm$0.059 & 13.259$\pm$0.056 & 12.241$\pm$0.058 \\
53431.4 & 18.306$\pm$0.061 & 16.840$\pm$0.060 & 14.905$\pm$0.066 & 13.282$\pm$0.059 & 12.258$\pm$0.059 \\
53433.3 & 18.122$\pm$0.058 & 16.677$\pm$0.058 & 14.749$\pm$0.063 & 13.153$\pm$0.056 & 12.201$\pm$0.056 \\
53434.3 & 18.038$\pm$0.060 & 16.622$\pm$0.056 & 14.712$\pm$0.057 & 13.152$\pm$0.054 & 12.220$\pm$0.057 \\
53435.3 & 18.377$\pm$0.060 & 16.885$\pm$0.062 & 14.902$\pm$0.062 & 13.285$\pm$0.056 & 12.344$\pm$0.057 \\
53438.3 & 17.934$\pm$0.059 & 16.511$\pm$0.060 & 14.590$\pm$0.063 & - & 12.204$\pm$0.061 \\
53439.3 & 17.738$\pm$0.055 & 16.386$\pm$0.057 & 14.539$\pm$0.056 & 13.053$\pm$0.055 & 12.209$\pm$0.055 \\
53439.3 & 17.738$\pm$0.055 & 16.386$\pm$0.057 & 14.539$\pm$0.056 & 13.053$\pm$0.055 & 12.209$\pm$0.055 \\
53442.3 & 17.738$\pm$0.057 & 16.360$\pm$0.057 & 14.485$\pm$0.060 & 12.986$\pm$0.056 & 12.176$\pm$0.057 \\
53443.3 & 18.027$\pm$0.058 & 16.605$\pm$0.055 & 14.680$\pm$0.061 & - & 12.343$\pm$0.056 \\
53444.3 & 17.576$\pm$0.056 & 16.232$\pm$0.057 & 14.371$\pm$0.057 & - & 12.116$\pm$0.057 \\
53445.3 & 17.636$\pm$0.061 & 16.254$\pm$0.060 & 14.379$\pm$0.060 & 12.887$\pm$0.055 & 12.056$\pm$0.058 \\
53446.3 & 17.473$\pm$0.060 & 16.107$\pm$0.060 & 14.233$\pm$0.057 & 12.790$\pm$0.055 & 11.960$\pm$0.056 \\
53447.3 & 17.230$\pm$0.056 & 15.917$\pm$0.060 & 14.057$\pm$0.066 & 12.634$\pm$0.056 & 11.808$\pm$0.056 \\
53448.3 & 17.409$\pm$0.059 & 16.077$\pm$0.057 & 14.244$\pm$0.070 & 12.790$\pm$0.057 & 11.992$\pm$0.059 \\
53449.4 & 16.923$\pm$0.059 & 15.623$\pm$0.061 & 13.816$\pm$0.076 & 12.397$\pm$0.058 & 11.573$\pm$0.058 \\
53450.3 & 16.769$\pm$0.058 & 15.467$\pm$0.060 & 13.664$\pm$0.065 & 12.242$\pm$0.057 & 11.438$\pm$0.057 \\
53452.3 & - & - & - & 12.183$\pm$0.058 & 11.367$\pm$0.060 \\
53453.3 & 16.765$\pm$0.057 & 15.462$\pm$0.063 & 13.688$\pm$0.068 & 12.269$\pm$0.056 & 11.455$\pm$0.059 \\
53458.3 & 16.613$\pm$0.057 & 15.300$\pm$0.057 & 13.501$\pm$0.059 & - & 11.252$\pm$0.056 \\
53459.3 & 16.481$\pm$0.057 & 15.165$\pm$0.058 & 13.372$\pm$0.061 & 11.951$\pm$0.054 & 11.143$\pm$0.055 \\
53460.3 & 16.389$\pm$0.057 & 15.083$\pm$0.057 & 13.289$\pm$0.060 & 11.928$\pm$0.056 & 11.093$\pm$0.056 \\
53461.3 & 16.587$\pm$0.057 & 15.295$\pm$0.065 & 13.512$\pm$0.061 & 12.101$\pm$0.057 & 11.269$\pm$0.057 \\
53462.3 & 16.334$\pm$0.058 & 15.043$\pm$0.061 & 13.231$\pm$0.070 & 11.834$\pm$0.055 & 10.992$\pm$0.058 \\
53463.3 & 16.388$\pm$0.055 & 15.076$\pm$0.062 & 13.281$\pm$0.070 & 11.862$\pm$0.055 & 11.028$\pm$0.056 \\
53464.3 & 16.498$\pm$0.056 & 15.194$\pm$0.062 & 13.386$\pm$0.068 & 11.979$\pm$0.055 & 11.131$\pm$0.059 \\
53465.3 & 16.384$\pm$0.055 & 15.094$\pm$0.057 & 13.287$\pm$0.061 & 11.899$\pm$0.054 & 11.056$\pm$0.055 \\
53466.3 & 16.510$\pm$0.054 & 15.194$\pm$0.058 & 13.392$\pm$0.059 & 11.976$\pm$0.054 & 11.124$\pm$0.055 \\
53467.3 & 16.400$\pm$0.053 & 15.088$\pm$0.053 & 13.267$\pm$0.056 & 11.852$\pm$0.053 & 11.031$\pm$0.053 \\
53468.3 & 16.389$\pm$0.058 & 15.084$\pm$0.056 & 13.276$\pm$0.059 & 11.872$\pm$0.055 & 11.061$\pm$0.057 \\
53469.3 & - & 15.324$\pm$0.062 & 13.513$\pm$0.067 & 12.104$\pm$0.056 & 11.265$\pm$0.056 \\
\hline
\end{tabular}
\end{table*}

\begin{table}
\caption{VLA flux densities of GRO~J1655-40. Calendar dates and MJD's denote mid-point of target observation. 1$\sigma$ errors are uncertainties on the fitted source parameters. The 3$\sigma$ upper-limit is determined from the image rms.
\label{table:radio}}
\centering
\begin{tabular}{llcc}
\hline
Date & MJD & Frequency & Flux density \\
 (UT) &  & (GHz) & (mJy\,beam$^{-1}$) \\
\hline
2005 Mar 18 & 53447.6 	& 1.425   & 6.4$\pm$0.4 \\
	    &           & 4.86   & 3.9$\pm$0.1 \\
            &   	& 8.46   & 2.5$\pm$0.1 \\
            &   	& 22.46  & 1.8$\pm$0.4 \\

2005 Mar 20 & 53449.5  	& 1.425   & 5.4$\pm$0.3 \\
       	    &   	& 4.86   & 2.4$\pm$0.1 \\
	    &   	& 8.46   & 1.91$\pm$0.09 \\

2005 Mar 22 & 53451.6  	& 1.425   & 3.0$\pm$0.3 \\
	    &   	& 4.86   & 1.4$\pm$0.1 \\
	    &   	& 8.46   & 1.14$\pm$0.08 \\

2005 Mar 23 & 53452.5  	& 1.425   & 2.3$\pm$0.3 \\
	    &   	& 4.86   & 1.0$\pm$0.1 \\
	    &  	        & 8.46   & 0.86$\pm$0.07 \\

2005 Mar 24 & 53453.6  	& 1.425   & 2.8$\pm$0.3 \\
	    &           & 4.86   & 1.4$\pm$0.2 \\
	    &   	& 8.46   & 0.86$\pm$0.09 \\

2005 Mar 31 & 53460.5  	& 1.425   & $\leq$0.81 \\
            &           & 4.86   & 0.4$\pm$0.1 \\
	    &   	& 8.46   & 0.25$\pm$0.05 \\

2005 Apr 05 & 53465.5  	& 1.425   & 0.9$\pm$0.3 \\
	    &           & 4.86   & 0.5$\pm$0.1 \\
	    &   	& 8.46   & 0.15$\pm$0.04 \\

2005 Apr 07 & 53467.4  	& 1.425   & 0.7$\pm$0.2 \\
	    &           & 4.86   & $\leq$0.56 \\

\hline

\end{tabular}
\end{table}

\begin{table*}
  \centering 
  \caption{\emph{eqpair} fit parameters}
  \label{table:eqp}
  \begin{tabular}{lccccccc}
   \hline
\# &kT &$l_{h}/l_{s}$$^{a}$ &$l_{nt}/l_{th}$$^{b}$ &$\tau^{c}$ &Ref. Fr.$^{d}$ &EW1$^{e}$ &EW2$^{e}$ \\
 &(keV) &  &  &  &  &(keV) &(keV) \\ \hline
1 &- & 10.94$\pm$3.10 & $<$0.16 & $<$0.32 & 0.36$\pm$0.15 & - & - \\
2 &0.30$\pm$0.02 &  7.57$\pm$2.27 & $<$0.08 & $<$0.18 & 0.72$\pm$0.14 & - & - \\
3 &0.26$\pm$0.08 & 13.32$\pm$3.27 & $<$0.50 & $<$0.50 & 0.29$\pm$0.12 & - & - \\
4 &0.31$\pm$0.04 & 16.69$\pm$2.88 & $<$0.22 & $<$0.53 & 0.09$\pm$0.06 & - & - \\
5 &0.28$\pm$0.05 & 12.77$\pm$2.78 & $<$0.50 & $<$0.41 & 0.34$\pm$0.12 & - & - \\
6 &0.32$\pm$0.08 & 16.56$\pm$2.10 & $>$0.82 & $<$0.83 & 0.15$\pm$0.11 & - & - \\
7 &0.31$\pm$0.06 & 18.09$\pm$5.04 & $<$0.31 & $<$0.65 & 0.19$\pm$0.09 & - & - \\
8 &0.31$\pm$0.12 & 23.20$\pm$6.31 & $<$0.88 & 0.64$\pm$0.63 & $<$0.12 & - & - \\
9 &0.34$\pm$0.04 & 18.73$\pm$1.52 & $<$0.12 & $<$0.38 & 0.07$\pm$0.05 & - & - \\
10 &0.30$\pm$0.04 & 17.89$\pm$1.40 & $<$0.44 & $<$0.57 & $<$0.03 & - & - \\
11 &0.30$\pm$0.03 & 17.01$\pm$1.46 & $<$0.18 & $<$0.66 & $<$0.14 & - & - \\
12 &0.31$\pm$0.02 & 14.93$\pm$4.23 & $<$0.10 & $<$0.24 & 0.11$\pm$0.04 & - & - \\
13 &0.29$\pm$0.04 & 17.15$\pm$1.55 & $<$0.33 & $<$0.44 & $<$0.10 & - & - \\
14 &0.45$\pm$0.21 & 17.38$\pm$1.58 &  & $<$1.12 & $<$0.13 & - & - \\
15 &0.22$\pm$0.04 & 17.89$\pm$0.92 & 0.55$\pm$0.20 & 1.77$\pm$0.24 & 0.21$\pm$0.04 & - & - \\
16 &0.31$\pm$0.04 & 19.46$\pm$2.12 & $<$0.94 & $<$0.59 & 0.22$\pm$0.05 & - & - \\
17 &$>$0.24 & 21.33$\pm$2.21 & $>$0.91 & $<$0.91 & 0.32$\pm$0.08 & - & - \\
18 &0.41$\pm$0.11 & 19.78$\pm$4.35 & 0.69$\pm$0.07 & 1.97$\pm$0.16 & 0.18$\pm$0.06 & - & - \\
19 &0.39$\pm$0.08 & 21.32$\pm$5.33 & $>$0.95 & 2.27$\pm$0.33 & 0.28$\pm$0.07 & - & - \\
20 &0.33$\pm$0.07 &  9.82$\pm$0.61 & $>$0.70 & 1.96$\pm$0.10 & 0.14$\pm$0.04 & - & - \\
21 &0.31$\pm$0.03 &  9.40$\pm$0.48 & 0.65$\pm$0.03 & 2.25$\pm$0.06 & 0.21$\pm$0.05 & - & - \\
22 &0.35$\pm$0.10 &  5.95$\pm$0.15 & 0.96$\pm$0.03 & 1.87$\pm$0.11 & 0.33$\pm$0.04 & - & - \\
23 &0.37$\pm$0.07 &  4.37$\pm$0.25 & $>$0.88 & 1.84$\pm$0.09 & 0.30$\pm$0.04 & - & - \\
24 &0.75$\pm$0.03 &  0.89$\pm$0.02 & $>$0.99 & 1.13$\pm$0.07 & 0.35$\pm$0.04 & - & - \\
25 &0.80$\pm$0.01 &  0.35$\pm$0.01 & $>$0.96 & 0.93$\pm$0.04 & 1.21$\pm$0.16 & -0.032 & - \\
26 &0.89$\pm$0.02 &  0.22$\pm$0.02 & $>$0.97 & 0.53$\pm$0.14 & 1.18$\pm$0.31 & - & - \\
27 &0.96$\pm$0.01 &  0.21$\pm$0.02 & $>$0.99 & 0.32$\pm$0.05 & 1.19$\pm$0.10 & - & - \\
28 &1.07$\pm$0.01 &  0.17$\pm$0.01 & $>$0.97 & $<$0.05 & 0.97$\pm$0.19 & - & - \\
29 &1.12$\pm$0.01 &  0.09$\pm$0.02 & $>$0.96 & $<$0.05 & 0.88$\pm$0.25 & -0.062 & -0.112 \\
30 &1.15$\pm$0.00 &  0.12$\pm$0.02 & 0.95$\pm$0.03 & $<$0.04 & $>$1.05 & -0.017 & - \\
31 &1.20$\pm$0.01 &  0.12$\pm$0.01 & 0.97$\pm$0.02 & $<$0.06 & 1.05$\pm$0.24 & -0.066 & -0.090 \\
32 &1.19$\pm$0.01 &  0.04$\pm$0.01 & 0.77$\pm$0.02 & 0.10$\pm$0.02 & $>$1.91 & -0.105 & -0.186 \\
33 &1.22$\pm$0.01 &  0.07$\pm$0.01 & 0.91$\pm$0.03 & $<$0.05 & $>$0.83 & -0.056 & -0.090 \\
34 &1.23$\pm$0.01 &  0.04$\pm$0.00 & 0.71$\pm$0.06 & 0.06$\pm$0.01 & 1.71$\pm$0.07 & -0.068 & -0.157 \\
35 &1.21$\pm$0.02 &  0.04$\pm$0.00 & 0.50$\pm$0.06 & 0.31$\pm$0.06 & $>$1.41 & -0.114 & -0.213 \\
36 &1.16$\pm$0.02 &  0.03$\pm$0.00 & 0.12$\pm$0.03 & $>$0.44 & $>$1.28 & -0.122 & -0.214 \\
37 &1.15$\pm$0.03 &  0.04$\pm$0.01 & 0.20$\pm$0.03 & $>$0.80 & $>$1.70 & -0.115 & -0.212 \\
38 &1.17$\pm$0.03 &  0.05$\pm$0.01 & 0.30$\pm$0.04 & $>$0.83 & $>$1.93 & -0.133 & -0.250 \\
39 &1.25$\pm$0.02 &  0.12$\pm$0.00 & 0.96$\pm$0.03 & 0.24$\pm$0.07 & 1.53$\pm$0.44 & -0.110 & -0.208 \\
40 &1.22$\pm$0.01 &  0.07$\pm$0.00 & 0.71$\pm$0.03 & 0.29$\pm$0.07 & $>$1.90 & -0.115 & -0.240 \\
41 &1.17$\pm$0.02 &  0.07$\pm$0.01 & 0.25$\pm$0.04 & $>$0.84 & $>$1.66 & -0.130 & -0.237 \\
42 &1.24$\pm$0.02 &  0.05$\pm$0.01 & 0.07$\pm$0.05 & 0.60$\pm$0.05 & 1.40$\pm$0.16 & -0.116 & -0.262 \\
43 &1.18$\pm$0.02 &  0.08$\pm$0.01 & 0.35$\pm$0.05 & $>$0.88 & $>$1.63 & -0.145 & -0.255 \\
\hline
\multicolumn{8}{l}{$^a$ Hard to soft compactness ratio}\\
\multicolumn{8}{l}{$^b$ Ratio of non-thermal electrons to thermal electrons}\\
\multicolumn{8}{l}{$^c$ Optical depth}\\
\multicolumn{8}{l}{$^d$ Reflection fraction}\\
\multicolumn{8}{l}{$^e$ EW1 and EW2 are equivalent widths of absorption features peaking at 6.7 keV and 8 keV, respectively.}
\end{tabular}
\end{table*}

% Don't change these lines
\bsp	% typesetting comment
\label{lastpage}

\end{document}